\documentstyle[epsf,12pt]{article}
\begin{document}

\catcode`@=11
\long\def\@caption#1[#2]#3{\par\addcontentsline{\csname
  ext@#1\endcsname}{#1}{\protect\numberline{\csname
  the#1\endcsname}{\ignorespaces #2}}\begingroup
    \small
    \@parboxrestore
    \@makecaption{\csname fnum@#1\endcsname}{\ignorespaces #3}\par
  \endgroup}
\catcode`@=12
\newcommand{\newc}{\newcommand}
\newc{\gsim}{\lower.7ex\hbox{$\;\stackrel{\textstyle>}{\sim}\;$}}
\newc{\lsim}{\lower.7ex\hbox{$\;\stackrel{\textstyle<}{\sim}\;$}}
\newc{\gev}{\,{\rm GeV}}
\newc{\mev}{\,{\rm MeV}}
\newc{\ev}{\,{\rm eV}}
\newc{\kev}{\,{\rm keV}}
\newc{\tev}{\,{\rm TeV}}
\newc{\mz}{m_Z}
\newc{\mpl}{M_{Pl}}
\newc{\chifc}{\chi_{{}_{\!F\!C}}}
\newc\order{{\cal O}}
\newc\CO{\order}
\newc\CL{{\cal L}}
\newc\CY{{\cal Y}}
\newc\CH{{\cal H}}
\newc\CM{{\cal M}}
\newc\CF{{\cal F}}
\newc\CD{{\cal D}}
\newc\CN{{\cal N}}
\newc{\eps}{\epsilon}
\newc{\re}{\mbox{Re}\,}
\newc{\im}{\mbox{Im}\,}
\newc{\invpb}{\,\mbox{pb}^{-1}}
\newc{\invfb}{\,\mbox{fb}^{-1}}
\newc{\yddiag}{{\bf D}}
\newc{\yddiagd}{{\bf D^\dagger}}
\newc{\yudiag}{{\bf U}}
\newc{\yudiagd}{{\bf U^\dagger}}
\newc{\yd}{{\bf Y_D}}
\newc{\ydd}{{\bf Y_D^\dagger}}
\newc{\yu}{{\bf Y_U}}
\newc{\yud}{{\bf Y_U^\dagger}}
\newc{\ckm}{{\bf V}}
\newc{\ckmd}{{\bf V^\dagger}}
\newc{\ckmz}{{\bf V^0}}
\newc{\ckmzd}{{\bf V^{0\dagger}}}
\newc{\X}{{\bf X}}
\newc{\bbbar}{B^0-\bar B^0}
\def\bra#1{\left\langle #1 \right|}
\def\ket#1{\left| #1 \right\rangle}
\newc{\sgn}{\mbox{sgn}\,}
\newc{\m}{{\bf m}}
\newc{\msusy}{M_{\rm SUSY}}
\newc{\munif}{M_{\rm unif}}
%
%
\def\NPB#1#2#3{Nucl. Phys. {\bf B#1} (19#2) #3}
\def\PLB#1#2#3{Phys. Lett. {\bf B#1} (19#2) #3}
\def\PLBold#1#2#3{Phys. Lett. {\bf#1B} (19#2) #3}
\def\PRD#1#2#3{Phys. Rev. {\bf D#1} (19#2) #3}
\def\PRL#1#2#3{Phys. Rev. Lett. {\bf#1} (19#2) #3}
\def\PRT#1#2#3{Phys. Rep. {\bf#1} (19#2) #3}
\def\ARAA#1#2#3{Ann. Rev. Astron. Astrophys. {\bf#1} (19#2) #3}
\def\ARNP#1#2#3{Ann. Rev. Nucl. Part. Sci. {\bf#1} (19#2) #3}
\def\MPL#1#2#3{Mod. Phys. Lett. {\bf #1} (19#2) #3}
\def\ZPC#1#2#3{Zeit. f\"ur Physik {\bf C#1} (19#2) #3}
\def\APJ#1#2#3{Ap. J. {\bf #1} (19#2) #3}
\def\AP#1#2#3{{Ann. Phys. } {\bf #1} (19#2) #3}
\def\RMP#1#2#3{{Rev. Mod. Phys. } {\bf #1} (19#2) #3}
\def\CMP#1#2#3{{Comm. Math. Phys. } {\bf #1} (19#2) #3}
\relax
%
%
%
\def\beq{\begin{equation}}
\def\eeq{\end{equation}}
\def\bea{\begin{eqnarray}}
\def\eea{\end{eqnarray}}
%
%
%
\newc{\ie}{{\it i.e.}}          \newc{\etal}{{\it et al.}}
\newc{\eg}{{\it e.g.}}          \newc{\etc}{{\it etc.}}
\newc{\cf}{{\it c.f.}}
\def\smuon{{\tilde\mu}}
\def\neut{{\tilde N}}
\def\char{{\tilde C}}
\def\bino{{\tilde B}}
\def\wino{{\tilde W}}
\def\higgsino{{\tilde H}}
\def\sneut{{\tilde\nu}}
\def\stau{{\tilde\tau}}
%
%
%
%
\def\slash#1{\rlap{$#1$}/} 
\def\Dsl{\,\raise.15ex\hbox{/}\mkern-13.5mu D} 
\def\delsl{\raise.15ex\hbox{/}\kern-.57em\partial}
\def\Ksl{\hbox{/\kern-.6000em\rm K}}
\def\Asl{\hbox{/\kern-.6500em \rm A}}
\def\Qsl{\hbox{/\kern-.6000em\rm Q}}
\def\gradsl{\hbox{/\kern-.6500em$\nabla$}}
%
%
%
\def\bar#1{\overline{#1}}
\def\vev#1{\left\langle #1 \right\rangle}
%

\begin{titlepage}
~~
\vskip 2cm
\begin{center}
{\large\bf Universal Extra Dimensions and Charged LKPs}
\vskip 1cm
{\normalsize\bf
Mark Byrne} \\
\vskip 0.5cm
{\it Department of Physics, University of Notre Dame\\
Notre Dame, IN~~46556, USA\\[0.1truecm]
}

\end{center}
\vskip .5cm

\begin{abstract}

We consider the possibility of charged stable Kaluza-Klein leptons in orbifold models in which all the Standard Model fields propagate (UED models). At tree level the masses of the 1st mode states are nearly identical and kinematics can prevent the decays of the 1st mode charged leptons.  Perturbative and non-perturbative affects alter this picture and allow for a range of possibilities. For single extra dimensional UED-type models with stable Kaluza-Klein charged leptons we calculate their production from the all-particle incident flux of cosmic rays and find that null results in direct searches for anomalously heavy water suggest a lower bound on the compactification scale: $1/R > 290 \gev$.

\end{abstract}

\end{titlepage}

\setcounter{footnote}{0}
\setcounter{page}{1}
\setcounter{section}{0}
\setcounter{subsection}{0}
\setcounter{subsubsection}{0}


There has been a renaissance in the theoretical investigation of extra dimensions
 focusing on realistic string-inspired constructions with calculable phenomenological predictions 
in upcoming collider searches and other terrestrial experiments.  This paper will focus on 
a specific class of models in which the extra dimensions are compactified on the $\tev ^{-1}$ scale, 
with a phenomenology presumably on the verge of being elucidated by the LHC or NLC.
We work within the framework of Universal Extra Dimensions (UED) \cite{Barbieri,Cheng1}
 in which the Standard Model fields and their Kaluza-Klein (KK) excitations propagate on an $R^4 \times S_1/Z_2$ bulk spacetime. 
 Constraints on the these models are generally weak relative to other extra dimensional models due to tree level KK number conservation \cite{Appelquist}.  In addition, beyond tree level there remains a ``KK parity" defined by the orbifold construction and subsequent assignment of the standard model fields, which implies a stable 1st mode KK excitation.

There have been a variety of analyses on universal extra dimensions to date including general collider phenomenology (\cite{Rizzo}, \cite{Dicus}, \cite{Macesanu}), 
Higgs physics (\cite{Appelquist2}, \cite{Petriello}), rare K and B decays \cite{Buras}, $b \rightarrow s \gamma$ \cite{Agashe}, $Z \rightarrow B \bar{B}$  \cite{Oliver}, and constraints from $B^{0} \bar{B^{0}}$ 
mixing \cite{Chakraverty}. There are additional studies advocating the possibility that the lightest Kaluza-Klein particle (LKP) may play the role of a viable dark matter candidate \cite{Chengdm}, \cite{ Servant}, \cite{Hooper}. These analyses and precision electroweak constraints \cite {Appelquist} suggest a compactification scale $ > 300 \gev$, although non-perturbative affects could conceivably lower this value (or raise it). 
 
In UEDs the orbifold and field assignment allow all the Standard Model (SM) fields to propagate in the bulk while remaining consistent with experimental constraints such as those listed above. However, the democracy is not really universal since loop corrections and possible non-perturbative effects introduce terms localized at the fixed points of the orbifold which are non-universal \cite{georgi2}, \cite{carena}. Of particular interest to extra dimensional phenomenology are corrections to the masses of the fields that propagate in the bulk due to renormalization of the fields at the fixed points \cite{chengms}.  That is, the orbifold fixed points (fixed under $y \rightarrow -y$) are unique and delta function terms arise perturbatively in the effective four dimensional Lagrangian if neglected at tree level. The inclusion of loop effects in the $S_1/Z_2$ orbifold breaks the democracy of the model and removes the degeneracies in the mass spectra at each level. The perturbative effects induce four dimensional operators localized on the branes which were absent initially and allow the even (odd) fields to mix with each other. The coefficients of these local operators from non-perturbative affects are incalculable. We will refer to the case in which they vanish at the UV cutoff as Minimal Universal Extra Dimensions (MUEDs) \cite{chengms}. We neglect gravitational affects for the remainder of the paper and assume the KK parity remains valid in the non-perturbative regime {\footnote{Ref.~{\cite{feng}} considers the case in which the first mode KK graviton is the LKP.}}.

This paper will consider the constraints on UEDs for the case in which the LKP is electrically charged (e.g., the 1st mode electron).  These constraints will not rely on the details of the cosmology in these models (see Ref.~{\cite{mohapatra}}). Rather, the constraints are based on cosmic ray production of these states in the upper atmosphere combined with searches for anomalously heavy water {\cite{dimopoulos}}. The key assumptions in the analysis are that the cosmic ray flux has been the same order of magnitude as it is now for $\sim 1$ billion years, and that the anomalously heavy water is uniformly distributed in the oceans and terrestrial samples over that time scale. The bounds derived are applicable to generic orbifold models with at least one stable electrically charged fermion in the bulk. For the remainder of the paper, lepton will refer only to the charged leptons (e, $\mu$, $\tau$).
  
In section I, we will briefly review the construction of universal extra dimensions.  Section II will discuss the degeneracies of the first level KK states and the prediction of charged stable states.  Section III considers the bulk and boundary loop corrections on the orbifold.  In the minimal scenario of Ref.~\cite{chengms} we calculate the conditions for a charged LKP and compute the lowest value of the compactification scale consistent with a neutral LKP (the 1st mode B field). We also consider long-lived states and find the regions of parameter space consistent with prompt decay inside the detector. For the more general case of a stable first mode lepton, we consider their production by cosmic rays and calculate the expected anomalous concentration of water in the oceans following Ref.~\cite{byrne}.        

\section{Section I: Fields and Orbifolds}

The field content of UEDs are the zero mode Standard Model fields and their KK excitations, which includes both chiralities for any fermion.  Since fermions are generally non-chiral in five dimensions, boundary conditions at the fixed points of the orbifold may be chosen so that the appropriate chiral zero mode fields are present and the unobserved chiral modes appear for $n \geq 1$. The extra dimension is wrapped on a circle of radius $2\pi R$, with the points $y \rightarrow -y$ of the extra dimension identified. Equivalently, the extra dimension is a line segment with fixed points of the $Z_2$ at the boundaries $ (0, \pi R)$ so that the fields propagate on ${\cal R}^4 \times [0, \pi R]$. The fields are either even or odd under the $Z_2$ and this parity fixes the stability of the first order KK modes implying a lightest Kaluza-Klein particle (LKP). Of course, if there are fields confined to the fixed points their interactions are not parity restricted and they can mediate transitions beween any KK levels.  

The effective 4D Lagrangian is specified by integration over the orbifold with the usual KK masses $\sim (1/R)$ from the $\Gamma^5 \partial_5$ terms, excluding electroweak symmetry breaking effects and Yukawa couplings. The details are given, for example, in Ref.~\cite{Petriello}. We limit the discussion to the zeroth and first mode KK spectrum for the remainder of the paper for the sake of relevance and simplicity. 

\section{Section II: UED Degeneracies}

Let $\{L,e\}$ denote the lepton doublets and singlets, $ \{Q,u,d \}$ the quark doublets and singlets and $\{\gamma, g\}$ the photon and gluon.  We note that there are, for example, two first mode electrons corresponding to the singlet and doublet states. Including Yukawas introduces additional mixing between the singlets and the doublets within a level but the mass splitting is typically small with the exception of the top quark states. The singlet/doublet mass matrix is as follows:

\[ \left( \begin{array}{ccc}
    n/R & m^{(0)} \\
    m^{(0)} & -n/R  \end{array} \right)\]

\noindent
In the above, $ m^{(0)}$ refers to the mass of the corresponding SM state. For the photon and Z the usual mass matrix is slightly modified:

\[ \left( \begin{array}{ccc}
    (\frac{n}{R})^2 + \frac{1}{4}g_1 ^2 v^2 & \frac{1}{4} g_1g_2 v^2 \\
     \frac{1}{4} g_1 g_2 v^2 &  (\frac{n}{R})^2 + \frac{1}{4} g_2 ^2 v^2 \end{array} \right)\]

Most states have negligible corrections from their original $\frac{1}{R}$ mass so that all the fermion masses are almost degenerate with the exception of the top quark. The photon is the LKP and there is insufficient phase space ($ < \frac{1}{2} \mev$) for the 1st mode singlet electrons and muons to decay to the minimal final states of a zero mode electron (muon) and the LKP photon. Since this holds for all $1/R$ values, charged relics are a generic prediction of UEDs assuming the boundary terms are absent. We consider their production by cosmic rays in a later section. The degeneracies can be broken by nonperturbative operators which break the KK parity eliminating the LKP or perturbatively through local operators at the fixed points which renormalize the couplings.  We do not consider operators which break the KK parity and forthwith discuss perturbative affects.

\section{Section III: Broken Degeneracies and Charged LKPs}

In this section we will consider both bulk and boundary loop corrections to the masses of the KK states. We limit the analysis to the loop corrections calculated in Ref.~\cite{chengms}. For the orbifold there are bulk loop corrections which alter the masses of each gauge field independent of KK level:

\begin{equation}
\delta\, (m_{gauge}^2) \propto  -\, \frac{g_n^2\,\zeta(3)}{16\pi^4}  
\left(\frac{1}{R}\right)^2.
\end{equation}

\noindent There are also boundary corrections from the orbifold construction which require infinite renormalization (\cite{georgi2}) and the introduction of an explicit ultraviolet cutoff ($\Lambda$): 

\begin{equation}
{\delta}\, m_{fermion} \propto m_n \sum_{k} \frac{g_k^2}{16\pi^2} \ln(\Lambda R)^2.  
\end{equation}

\begin{equation}
{\delta}\, (m_{gauge_n}^2) \propto m_n^2 \frac{g_n^2}{16\pi^2} \ln(\Lambda R)^2.
\end{equation}    

\noindent For the fermion corrections the sum is over the relevant gauge fields and the proportionality constants are order one. 

With these corrections to the masses of the KK states we can examine the first mode mass spectrum of all the Standard Model KK states as the parameters $R$ and $\Lambda$ are varied.  For most of the parameter space, the KK photon is the lightest mode.  However, it is possible that the first mode singlet electron is the LKP or it is the next-to-lightest state and the near-degeneracy with the first mode photon prevents the decay from occuring. There is also the possibility of a near-degeneracy which allows the decay to proceed but with a sufficiently long lifetime that the state appears stable from a collider perspective. We are interested in finding the regions of parameter space which predict either a charged LKP or a long-lived charged state which appears to be stable since it decays outside the detector.

\begin{figure}
\centering
\epsfysize=3.2in
\hspace*{0in}
\epsffile{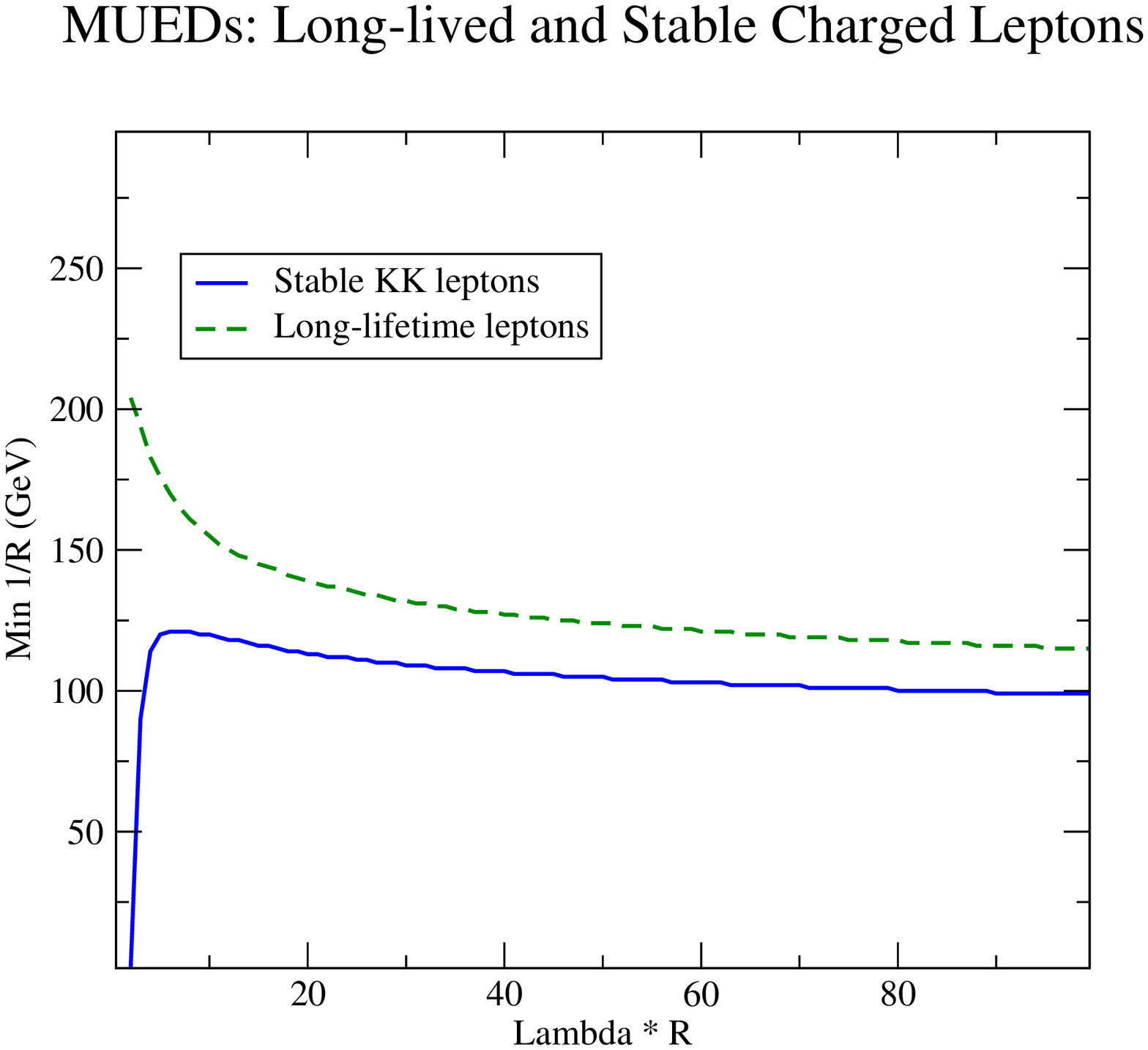}
\caption{Assuming the boundary corrections of Ref.~\cite{chengms}, the minimum value of the compactification scale consistent with the absence of stable (solid line) or long-lived (dashed) charged leptons vs. the effective number of KK modes below the cutoff $\Lambda$. Above the solid line the $\gamma^1$ is the LKP. Between the solid and dashed line there is at least one massive long-lived charged lepton.}     
\label{fig1}
\end{figure}

Generally there are three possibilities in this MUED scenario depending on the mass spectrum for the 1st mode states: a lepton, or leptons are the LKP; the photon is the LKP but the lepton-photon mass splitting is less than the SM lepton mass; the photon is the LKP and the lepton-photon mass splitting is larger than the SM lepton mass. For the last case the decays are prompt and the phenomenology has been considered in some detail \cite{chengms2}. For the case where the LKP is a lepton we will examine the implied constraints on the compactification scale from cosmic rays. If the KK lepton-photon mass splitting is less than the SM lepton mass we determine the lifetime of the KK lepton as a function of the mass splitting and classify the state as long-lived if the lifetime exceeds $10^{-6}$ seconds. \footnote{Criteria from collider searches for heavy leptons \cite{opal}.} 

	For the long-lived states, the usual decay $lepton^{(1)} \rightarrow lepton^{(0)} + \gamma^{(1)}$ is absent kinematically.  The decay can proceed via a virtual SM lepton and virtual SM W with four final states. For example, $\mu^{(1)} \rightarrow  \gamma^{(1)} + \nu_{\mu}^{(0)} + \nu_{e} ^{(0)} + e^{(0)}$.  In this case, the KK muon lifetime is greater than the lifetime of the SM muon ($2.2 \times 10^{-6} s$) which satisfies our criteria for a massive lepton escaping the detector so that no further calculations are necessary to determine the maximum mass splitting necessary for stable 1st mode leptons. The analysis proceeds similarly for the taus, where in this case the lifetime is sufficiently long if the 1st mode photon-tau mass difference is of the order of the SM muon mass or smaller. We examine the parameter space associated with these long-lived states. 

%
%

In Figure 1 we determine the minimum value of the compactification scale consistent with no stable KK leptons (solid line), and no long-lived muons and taus (dashed-line) as a function of $\Lambda R$.  More simply stated, below the dashed line there are long-lived leptons; below the solid line the leptons are stable. The absence of charged LKPs in the low $\Lambda R$ region is due to large negative bulk contributions to the $B^1$ field, while the boundary corrections fall off with log $\Lambda R$.  Direct bounds on the mass of heavy singly charged leptons are $\sim 100 \gev$ \cite{opal} and we find MUEDs predict stable leptons up to $150 \gev (\Lambda R = 10)$.  We consider bounds on the compactification scale from cosmic ray production for the stable KK leptons in the next section.  These bounds are also applicable to the more general case in which the localized boundary terms are arbitrary and there is the possibility of a charged LKP for all values of $\frac{1}{R}$.

\section{Cosmic Ray Production of KK states}
We present a quick summary of cosmic ray production of stable KK leptons and estimate the bounds derivable from null results in searches for anomalously heavy water.  The details can be found in Ref.~\cite{byrne}. We compute the QCD constituent cross sections for the KK gluons and quarks using the squared matrix elements of Ref.~\cite{Macesanu}. The QCD cross sections are higher than the direct QED cross sections by a factor of $\frac{\alpha_s}{\alpha}$ so we neglect direct production of the lepton LKP.  Since we are interested in the regime where the LKP is a lepton, the KK quarks and gluons will decay into a minimum of one positively charged LKP in the final decay products.  These will bind an electron becoming chemically indistinguishable from hydrogen, except its mass $\sim 1/R$ rather than $1 \gev$; the hydrogen typically forms heavy water {\cite{dimopoulos}} (assuming only half the anomalously heavy hydrogen molecules form heavy water changes the final bound by less than $5 \gev$). Smith et.~al.~\cite{smith} has searched for these anomalously heavy water molecules and we indicate the bounds on the compactification scale from these searches in Figure 2. The dependence on $\Lambda$ is from converting the bound on the gluon and quark masses to the LKP mass for a given value of $1/R$. For $\Lambda R = 10$ we find that $1/R > 245 \gev$ and for $\Lambda R =100, 1/R > 220 \gev$.  This appears to rule out the regions of parameter space found in the previous section predicting charged LKPs. There is the possibility that non-perturbative effects alter the coefficients of the boundary terms such that a lepton is the LKP or it remains stable by kinematics. These corrections do not need to be large. The corrections to the masses of the singlet leptons in the minimal model were ${\cal O} (\frac{g_1 ^2}{16 \pi ^2}) \sim 10^{-3}$. It is easy to see that any appreciable negative corrections (1 percent or more) to the masses of the fermions from unknown UV effects at the fixed points will easily push their masses below the 1st mode photon implying charged relics in the theory. In these cases the cosmic ray calculations suggest $1/R > 290 \gev$, comparable to electroweak precision bounds.  
  
\begin{figure}
\centering
\epsfysize=3.2in
\hspace*{0in}
\epsffile{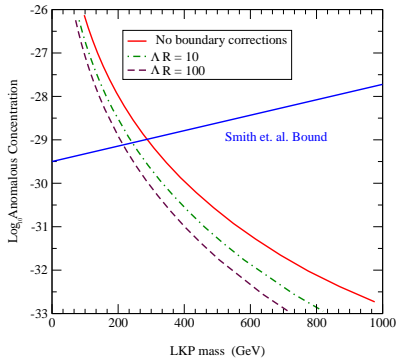}
\caption{For the case of a charged LKP, the anomalous concentration of heavy water in the oceans vs. the LKP mass alongside the bound of Ref.~{\cite{smith}}.}  
\label{fig2}
\end{figure}

\section{Conclusions}
In extra dimensional theories the level one KK states are typically degenerate in mass at tree level.  If the compactification scenario introduces a discrete parity which prevents the decay of the lightest first mode states then there exists the possibility of stable charged KK states.  The corrections to the tree level masses are of primary importance in determining the decay pattern for the states in the theory and depend on both loop effects and non-perturbative effects.  In this paper we have considered an $S_1/Z_2$ model in which all the standard model fields propagate and place bounds from cosmic ray production assuming the LKP is electrically charged. Whether the LKP is charged or not depends on the unknown behavior of the theory above the cutoff.  We consider a minimal model with only radiative corrections at the boundary of the $S_1/Z_2$ orbifold which alters the mass spectrum and find that the model avoids charged relics above $\sim 120 \gev$ and long-lived leptons above $\sim 200 \gev$.  If there are additional corrections to the boundary operators and these corrections imply electrically charged LKPs then cosmic rays place a lower bound on the compactification scale of $290 \gev$ independent of the cosmology of these models.

\section*{Acknowledgments}
I would like to thank C.~ Kolda, D.~Benson, S.~Mouslopoulous, D.~Bennett, and J.~Losecco for useful comments. I am grateful for the hospitality of the T-8 division at Los Alamos National Lab (LANL) where a portion of this work was completed. M.B. is supported in part by a fellowship from the Center for Applied Mathematics (CAM) at the University of Notre Dame.

\end{document}